\documentclass[12pt]{article}
\makeatletter

\@addtoreset{equation}{section}
\makeatother
\usepackage{amsfonts}
\usepackage{eucal}

\renewcommand{\Im}{{\rm Im}\,}

\begin{document}

\title{\bf When does ``Time" become ``Observable" ?}

\author{Tadashi Fujimoto\\
Department of philosophy, Ryukoku University\\
Kyoto 600-8268\\
Japan\\
E-mail: fujimoto-t@let.ryukoku.ac.jp}
\date{March. 2014}
\maketitle

\section*{Summary}
We investigate some spectral properties of time operators  which are obtained through Canonical Commutation Relation (CCR) and Positive Operator Valued Measure (POVM) of  quantum physics. In addition, we re-interpret the spectral properties of time operators from the standpoint of the Copenhagen Interpretation, especially, of W.Heisenberg.   

\medskip

\noindent
{\bf Keywords}:  Hamiltonian, Time operator, Positive Operator Valued Measure, Spectrum

\section{Introduction}
In our (physical) world, we can recognize ``time" through spatial movement (for example, clock, movement of the sun or moon).  But from where does come this ``time"?  we can recite many philosopher who had considered `` time" (cf, St.Augustine, I.Kant, H.Bergson, M.Heidegger, et al.)
Historically, ``time" have been the subject of considerable interest. From the viewpoint of mathematics  and physics, especially operator theory,  what characteristics does ``time"  have?  
According to the Copenhagen Interpretation, we cannot directly know ``the quantum world itself", but at least we can know it through observation.  Is there a reason or a cause of our  phenomenal time?  One of the keys to this issue (mystery?) is in my opinion the ``time operator".

The purpose of this paper is to present a new philosophical interpretation with the historical  investigation in mind. In particular, the background knowledge required for this paper is the concept of the Copenhagen Interpretation and the characters on quantum physics.    

In the next section and section 3, we review main past studies on the time operator and categorize the existing methods so far proposed. Section 4 presents a new framework concerning the time operator. And in section 5, we consider philosophical interpretations.

\section{Three types of time in quantum physics}
We can classify the time in quantum physics [7].

(A)  external time: is relevant to experiment.

(B)  observation time: is relevant to observation (detector).

(C)  internal time: is relevant to the object itself.

\medskip
According to W.Heisenberg based on the Copenhagen Interpretation，here I generalize these definition (A), (B), (C) respectively as follows.

\medskip

(A)　is phenomenal time (classical time).

(B)　is the time of observation.

(C)　is non-phenomenal time (quantum time).

\medskip

(B) is the so-called bridge between (A) and (C).

Using terms of Aristotle(BC.384-322), (A) is in Energeia (Wirklichkeit) and (C) is in Dynamis (M\"{o}glichkeit).  Therefore we can only recognize ``the time of (C)" through the procedure (B) [9, 15].

It was well known that when we try to measure  the energy of particles in a non-steady (non-stationary) state, the experimental values are in dispersion. So the uncertainty between the energy-measurement and the time of observation was considered. Heisenberg analyzed the experiment of ``Stern-Gerlach" and he showed that  $\delta${E}$\delta${T}$\sim{h}$, i.e. the time-energy uncertainty relation

$$\Delta{T}\Delta{E}{\ge}\frac{1}{2}\hbar,$$
where $\Delta$ is the standard deviation [10, 11].

Next, Aharonov  and Bohm constructed an operator with dimension of time. This operator is named ``Aharonov-Bohm time operator"[5]. 

$$T=\frac{1}{2m}(Q{P^{-1}}+{P^{-1}}Q).$$

$Q$ is position operator, $P$ is momentum operator.   

It is conjugate to the free particle Hamiltonian $H=\frac{P^2}{2m}$.  i.e.  Canonical Commutation Relations (CCR),  $[T, H]=i$. ($\hbar=1$).
Is $T$ a physical quantity? Is the spectrum is real?  In other words, is $T$  (essentially) self-adjoint? 
W.E.Pauli answered ``No" to the above  question. 

 This fact is easy to see from ``Von Neumann uniqueness theorem"\footnote{ Theorem ``Von Neumann uniqueness":
 
Let q and p be on the Hilbert space $ L ^ {2} (\mathbf {R}) $ defined by 
$q:=M_x$, $p:=-iD_x$, where $M_x$ is the multiplication operator by variable $x\in{\mathbf{R}}$, and $D_x$ is the generalized differential operator in $x$. Then a self-adjoint pair $(q,p)$ satisfies the condition of the Weyl representation of CCR, i,e.  $$e^{isp}e^{itq}=e^{-ist}e^{itp}e^{isq}.$$

$(q,p)$ is called the Schr\"{o}dinger representation of CCR. 

Let $\mathcal{H}$ be separable Hilbert space and $({Q, P)}$ satisfies the condition of the Weyl representation of CCR. Then, if $(Q,P)$ is irreducible, there exists a unitary operator $U$: $\mathcal{H}{\to}L ^ {2} (\mathbf {R})$ s.t, $UQU^{-1}=q$, $UPU^{-1}=p$. 

The spectrum is the following: $\sigma{(Q)}=\sigma(q)=\mathbf{R}$, $\sigma(P)=\sigma(p)=\mathbf{R}$

Therefore, if $(T, H)$ were a self-adjoint pair, because $(T,H)$ is a Weyl representation,  $\sigma{(T)}=\sigma(q)=\mathbf{R}$, $\sigma(H)=\sigma(p)=\mathbf{R}$. } i.e. that there is no self-adjoint operator which satisfies CCR with the self-adjoint operator (Hamiltonian) bounded from below (or above) [10, 12].
      
But, Pauli only  indicated that for a semi-bounded self-adjoint Hamiltonian, there is no self-adjoint time operator. Nowadays we understand some types of time operators. Based on the methods of mathematical operator theory, there is no need to restrict within self-adjoint operator. We can study the time operator from this view point. In the next section, we overlook this fact. 

\section{Classification of time operators according to CCR}
The classification  which is basically introduced by A. Arai [2, 4] will be described in details below. The Aharonov-Bohm type is contained in these classification. 
 
Now, we always suppose that Hamiltonian $H$ is (essentially) self-adjoint (of course, we need to prove whether $H$ is self-adjoint or not.). If the time operator $T$ is self-adjoint and $H$ is unbounded self-adjoint, the Weyl representation exist. But here we would like to classify time operators for general $H$.   

If we allow for $T$ to be a symmetric operator, we can recognize several types of $T$.
In the following, we introduce  3-types of time operators.

\subsection{Definition ``Weyl representation"  type (1) }

As we have already seen, Paul denied this type of time operator from a viewpoint of physical meaning. But in a mathematical meaning, we regard this type as a kind of time operator. Then a self-adjoint pair $(T,H)$ satisfies the condition of the Weyl representation of CCR. i.e. 
$$e^{isH}e^{itT}=e^{-ist}e^{itH}e^{isT.}$$
Then $(T,H)$ is one of the Schr\"{o}dinger representations of CCR.  And these spectrum are  $\sigma{(T)}=\mathbf{R}$, $\sigma(H)=\mathbf{R}$. 

For example, Let be $H$ as follows: $H=aP$ (``$a$" is a real number). If we  regard  $T=\frac{Q}{a}$ as a time operator, then we can get it and this time operator is self-adjoint. 
Or, we can illustrate this case with a freely falling particle [7]. Let be $H$ as follows: $H=\frac{P^2}{2m}-mgQ$ ($g$ is the gravity acceleration). Then we get  $T=\frac{1}{mg}P$ as a self-adjoint time operator. 
 
\medskip

\subsection{Definition ``weak Weyl representation"  type (2)}

 A (closed) symmetric operator $T$ on $\mathcal{H}$ is called a ``weak Weyl representation"-type time operator [14, 16] w.r.t. a self-adjoint operator $H$ on $\mathcal{H}$  if, for all $t{\in}\mathbf{R}$, $e^{-itH}D(T)\subset{D(T)}$ and for all $\Psi\in{D(T)}$

$$Te^{-itH}\Psi=e^{-itH}(T-t)\Psi.$$

For example, the Aharonov-Bohm time operator is one of the type (2) operator w.r.t. $H=\frac{P^2}{2m}$. And the relativistic time operator $T$  is in the type(1) w.r.t the free Hamiltonian  for a relativistic quantum particle $H=\sqrt{-{\Delta}+m^{2}}$. (m$\ge${0}) : $T=HP^{-1}Q+QP^{-1}H$ (on a dense domain, using Fourier transformation). Further, we can get a time operator w.r.t. the Dirac type Hamiltonian and we can generalize the type (2) in Fock spaces. And if in the above, each Hamiltonian has a symmetric potential $V$ (i.e. $H+V$), we can get a time operator under some conditions using perturbation theory.

In addition, the following theorems concerning the time operator of the type (2) are known [2, 4, 13, 14, 16]:

\medskip

[Theorem (a)]

Let $H$ be a self-adjoint and semi-bounded operator on $\mathcal{H}$. Then there is no self-adjoint time operator $T$ of the type (2) w.r.t. $H$ which can be essentially self-adjoint. Therefore, in the case when $H$ is semi-bounded, $T$ is not observable (i.e, $\sigma(T) $ is not in $\mathbf{R}$). 

In other words, if $T$ is essentially self-adjoint, then $H$ and $T$ are not semi-bounded and

$$\sigma(H)=\sigma(T)=\mathbf{R}.$$
 
\medskip

[Theorem(b)]

 Suppose that a self-adjoint operator $H$ has a time operator $T$ of the type (2) which is conjugate to $H$. Then $H$ has only absolutely continuous spectrum. i.e. $\sigma(H)=\sigma_{ac}(H)$

\medskip

[Theorem(c)]

Let $H$ be a self-adjoint operator on $\mathcal{H}$ and let $T$ be a time operator of the the type (2) w.r.t $H$. Then the following hold:

(i) if $H$ is bounded from below, then $\sigma(T)$ is either $\mathbf{C}$ or $\overline{\{z\in\mathbf{C}| \Im{z}>{0}\}};$

(ii) if $H$ is bounded from above, then $\sigma(T)$ is either $\mathbf{C}$ or $\overline{\{z\in\mathbf{C}| \Im{z}<{0}\}};$

(iii) if $H$ is bounded, then $\sigma(T)=\mathbf{C}.$

\medskip

\subsection{Definition  ``normal "-type (3)}

 A (closed) symmetric operator $T$ on $\mathcal{H}$ is called a ``normal"-type time operator [3] w.r.t. a self-adjoint operator $H$ on $\mathcal{H}$, if there is subspace $\mathcal{D}\neq\{0\}$, $\mathcal{D}{\subset}D(TH){\cap}D(HT)$, for all $\Psi, \Phi\in\mathcal{D}$.

$$[T,H]\Psi=i\Psi.$$ 

Let $H$ be self-adjont on $\mathcal{H}$ with purely discrete spectrum: $\sigma(H)=\sigma_{p}(H)=\{E_{n}\}^{\infty}_{n=1}$ ($E_{n}<E_{n+1}$, $\lim{E}_{n \to \infty}=\infty$)
and the multiplicity of each $E_{n}$ being one and $\Sigma^{\infty}_{n=1}\frac{1}{E^{2}_{n}}<\infty$. Let $e_{n}$ be the normalized eigenvector of $H$ with eigenvalue $E_{n}$: $He_{n}=E_{n}e_{n}$ and $\|e_{n}\|=1$. Then we can define a time operator $T$ on $\mathcal{H}$:

$D(T):=\mathcal{L}(\{e_{k}-e_{l}| k, l \in\mathbf{N}\})$, for all $\Psi\in{D(T)},$

$T\Psi:=\Sigma_{n\neq{m}}\frac{i}{E_{n}-E_{m}}{\langle}e_{m}, \Psi{\rangle}e_{n}$.

\medskip

This time operator is very interesting and is called ``Galapon-time operator" and it belongs to the type (3), but this time operator does not belong to type (1), (2). And if $inf_{\{n,m{\in}\mathbf{N}\}}({E_n}-{E_m})\neq{0}$, this time operator is essentially self-adjoint and bounded\footnote{In the case of $inf_{\{n,m{\in}\mathbf{N}\}}({E_n}-{E_m})={0}$, $T$ is unbounded and we do not known anything about $T$ yet.}. 
Here, we look at the example of Galapon-time operator associated with  harmonic oscillator [9]:

Let $E_n=\omega(n+\frac{1}{2})$, $n\in\{0\}\cup\mathbf{N}$ with a constant $\omega>0$.
In this case $T$ is a bounded self-adjoint operator with $D(T)=\mathcal{H}$, and $T\Psi=\frac{i}{\omega}\Sigma_{n=1}(\Sigma_{n\neq{m}}\frac{{\langle}e_{m}, \Psi\rangle}{n-m})e_{n}$
, $\Psi\in\mathcal{H}$.

We can prove : $\sigma(T)=[\frac{-\pi}{\omega}, \frac{\pi}{\omega}]$.
As is well known, in the context of quantum physics, the sequence $\{\omega(n+1/2)\}_{n=1}$ appears as the spectrum of the one-dimensional quantum harmonic oscillator $H:=\frac{P^2}{2m}+\frac{1}{2}m{\omega^2}Q^2$  with mass $m>0$

\subsection{Time operator and symmetric operator}
From these results, we need to mention the relation between the time operator and the symmetric operator. In mathematical sens, the type(1) is a interesting model, but from physical view, we cannot approve this model as realistic one because of the unboundedness of Hamiltonian. The famous time operator  i.e. the Aharonov-Bohm time operator is in the type (2), and in this type, Hamiltonian is generally bounded, so in the time operator of this type a realistic model is shown. But the spectrum of the type (2) is not real number. Therefore we cannot regard this type time operator as the realistic observation undoubtedly. The time operator of the type (3), especially, associated with harmonic oscillator Hamiltonian, we can associate the movement of some clocks or watches on ``time" judging from the spectrum. In this sense, the type (3) provides  also realistic model in strong sense. But, Is the values of ``time" observable bounded? In generally speaking, we recognize or image ``time" as like one-dimensional picture represented in the parameter $t\in\mathbf{R}$. Of course, the movement of clock is periodic infinite, but we need to distinguish between this movement and the representation of ``time" of which we ordinarily are conscious. Because the events on our  world never exactly repeat. On this question we cannot answer here because we need to mention the irreversibility on ``time" concerning the thermodynamics.

And hence we re-interpret the relationship between the above three types according the Copenhagen Interpretation. Before re-interpretation, it is need to refer the method of Positive operator valued measure on the time operator.

\section{Positive Operator Valued Measure and time operator }
These above methods are based on CCR representations : $[T ,H]=i$. But this approach does not allow us to interpret difference between the time operator $T$ and the parameter $t$. Previous investigation of the time operator have tended to neglect this difference. K.Fredenghagen, R.Burunetti and P.Bush et al. have proposed that the problem of defining the time operator as using Positive Operator Valued Measure (POVM).  Therefore we here look at the method of POVM.  We can regard POVM as generalized Spectral Measures (SPM). First, the definition of POVM is shown below.

\medskip

[Definition:  POVM ] [17]:

Let $\mathcal{B}_{\Omega}$ be Borel-$\sigma$-field of sets on a compact Hausdorff space and $\mathcal{B}(\mathcal{H})_{+}:=\{A{\in}\mathcal{B}(\mathcal{H}) | A\ge0\}$. Then if a mapping $F$:  $\mathcal{B}_{\Omega}\to\mathcal{B}(\mathcal{H})_{+}$ satisfies the following conditions, F is called a POVM on $\Omega$.

\begin{itemize}
\item $F(\Omega)=I$, $F(\emptyset)=0$.

\item $F(S)=F(S)^{*}$  for all $S\in\mathcal{B}_{\Omega}$.

\item For mutually disjoint element of $\{S_{n}\}^{\infty}_{n-1}\subset\mathcal{B}_{\Omega}$, $F(\cup^{\infty}_{n=1}S_{n})=\Sigma^{\infty}_{n=1}F(S_n)$, (strong-convergence).

\end{itemize}

When $F$ is POVM with $F=F^{n}$, $F$ is a SPM.  As we know, the spectral theorem is valid when an operator corresponding to spectral measures is self-adjoint, bounded-normal and unitary operator. Namely,

$\langle\psi, T\phi{\rangle}=\int_{\mathbf{R}}{\lambda} d\langle\psi , F(\lambda)\phi\rangle$,  $\psi\in\mathcal{H}$, $\phi\in{D(T)}$, 

where $D(T):=\{\phi\in\mathcal{H}| \int_{\mathbf{R}}{\lambda^{2}} d\langle\phi , F(\lambda)\phi\rangle<\infty\}$.

But a POVM  only corresponds to an (but not unique) unbounded symmetric operator. 
\medskip

When we make up POVM from SPM (or SPM from POVM) Naimark' Dilation theory  is needed as follows [1]:

\medskip

[Theorem: M. Naimark]

 For all POVM $F$ on $\mathcal{K}$, there is a Hilbert space $\mathcal{H}$ which contains $\mathcal{K}$ and a SPM $E$ which satisfies the following:

$PE(S)P=F(S)$,   for all $S\in\mathcal{B}_{\Omega}$, where $P$ is projection : $\mathcal{H}\to\mathcal{K}$.

\medskip

In general, $F$ is not a SPM, so there is no theorem like the spectral theorem. Therefore, $F$ is not unique, so there is no one-to-one correspondence between a POMV and a symmetric operator $T$.  Using POVM, we can get the Aharonov-Bohm time operatora  as follows:

 Let H be a free particle Hamiltonian $H=\frac{p^2}{2m}$. Then we can define a POVM. 
$$F(S)=\frac{\sqrt{PQ}}{2\pi{m}}{\int}e^{it\frac{P^{2}-Q^{2}}{2m}}dt.$$

 Then, when we calculate the integrals of the first moment with this POVM, then $T$ is Aharanov-Bohm type [6] .

Of same Aharanov-Bohm type, there is another POVM $F$: [7]

 $$F(S)=\frac{1}{2\pi}\int_{S}{dt}\left(\biggl|\int_{0}^{\infty}dP\sqrt{\frac{P}{m}}e^{\frac{itP^{2}}{m}}\biggr|^{2}+\biggl|\int^{0}_{-\infty}dP\sqrt{\frac{-P}{m}}e^{\frac{itP^{2}}{m}}\biggr|^{2}\right).$$

We can here look at the example that POVM  coincide with SPM [7]. This model has already seen as the type (1) in the previous section.  

Let be H as follows: $H=\frac{P^2}{2m}-mgQ$ (g is the gravity acceleration). Then we get  $T=\int{\lambda}dF(-mg\lambda)=\frac{1}{mg}P$ as a self-adjoint time operator (freely falling particle).  

\medskip
And, as another example,

Let $H$ be a self-adjoint harmonic oscillator: $H=\frac{1}{2}(P^{2}+Q^{2})$, ($m=\hbar=1$).
Introduce the annihilation operator $A=\frac{1}{2}(Q+iP)$, which gives the number operator $N=A^{*}A$, then $$H=N+\frac{1}{2}I$$ (one-dimensional quantum harmonic oscillator).

Then, for $t\in[0, 2\pi]$, we define POVM as follows: mod $2\pi$,
$$F(S)=\Sigma_{n,m\ge0}\frac{1}{2\pi}{\int}e^{i(n-m)t}dt|n\rangle{\langle}m|,$$

$$T=\int^{2\pi}_{0}\lambda{dF}(\lambda)=\Sigma_{m\neq{n}\ge0}\frac{1}{i(n-m)}|n\rangle{\langle}m|+\pi{I.}$$

\medskip

These example represent type (1) and (3). 

\medskip

The advantage of using POVM is follows: we can recognize explicitly the difference between the self-adjoint operator and the symmetric operator in the context of measure theory.  In fact, Fredenhagen and Brunetti, using thier approach, make the relation between the time parameter $t$ of the Sch\"{o}dinger equation and  the Aharanov-Bohm time operator with Naimark dilation theorem. When we consider a time operator to be a symmetric operator, we recognize the time operator on some restricted Hilbert spaces concerning some measures.

\section{Philosophical interpretation on time operator }
We cannot recognize quantum time itself. Only we can understand it through macro-time and observation-time. So according to the Copenhagen Interpretation, the physical phenomenon (macro-time) is the only important object. If this were true, we cannot  reach quantum time (micro-time) and there is no existence of quantum world i.e.we will fall in ``Agnosticism". Of course it is impossible to see quantum time directly.  But, I think it is indirectly possible. When we consider something in the micro-world, we cannot help using  the mathematical structure.  

As is known, some symmetric operators play important roles in the context of quantum physics. For instant, in particle physics, the creation and annihilation operator are symmetric operators. we cannot the event which a particle create or annihilate in quantum world directly, only we can know is the result of the interaction of some particles. But If we understand the interaction of particles,  the creation and annihilation operator are essential tool. In the same way, when the spectrum of a time operator is not to be real number, there is some justifications for treating the time operator as the observation. First, because we have already know that some time operators have the spectrum on real axis, so it is possible to explain the position of the type (2)-time operator concerning the region of Hilbert spaces. Specifically, the type (2) is consider the time operator as  ``generalized observable" or ``pre-observable'' on subspaces of a enlarged Hilbert spaces. Here ``generalized", ``pre" mean as follows: that in some physical theory we use no ``real time" i.e. we use ``complex number time", for example, in the case of  Wigner measure theory, we construct analytic continuation. At that time ``the real time" $t$ is replace with ``the complex number time". As another example, we can illustrate the theory of Hawking's singularity. Therefore in view of these examples, there is no reason to exclude the symmetric time operator from the observable in general. Rather the symmetric time operator works implicitly when we try to  measure the object itself  in the quantum world.

\section{Conclusion}
When does ``Time" become ``Observable"?  Here is a tentative explanation for this: ``time" does not always explicitly appear in the feature of ordinary value of observation. But ``time" appears, as like the type (3), associated with some type of  Hamiltonians. At same time, in this case, POVM change into SPM. Therefore we have to grasp this subject from a broader perspective.
  
 We understand that the time operator has a close relation to energy. Namely, $[T,H]=i$. Therefore, when we think about ``time" in (quantum) physics, we have to consider the relation with the energy operator. In addition, using POVM, we understand the relation between the self-adjoint time operator and the symmetric time operator. 

But in general we cannot use CCR $[T,H]=i$ in curved space time, because there is no Hamiltonian on it. In this case, what kind of methods can we use? This problem is very difficult and interesting.

\section*{Acknowledgment}
I would like to express my sincere appreciation and gratitude to Prof. Klaus Fredenhagen and Dr.Thomas P Hack for their help during my staying (in academic year 2011) at the University of Hamburg (Desy). This study was supported by a research grant from the Research office of  Ryukoku University (Kyoto in Japan) .

\end{document}